\documentclass[12pt]{iopart}

\usepackage{iopams}
\usepackage{amssymb}
\usepackage{graphicx}

\date{today}

\newcommand{\beq}{\begin{eqnarray}}
\newcommand{\eeq}{\end{eqnarray}}
\newcommand{\nn}{\nonumber}

\begin{document}

\begin{flushright}
DCP-10-04
\end{flushright}

\title[Quantum charged rigid membrane]{Quantum charged rigid membrane}
\author{Ruben Cordero$^1$, Alberto Molgado$^{2,3}$\ and Efrain 
Rojas$^{4,5}$}
\address{$^1$Departamento de F\'isica, Escuela Superior de 
F\'\i sica y Matem\'aticas del I.P.N., 
Unidad Adolfo L\'opez Mateos,
Edificio 9, 07738 M\'exico, D.F., MEXICO }
\address{$^2$Unidad Acad\'emica de F\'isica, Universidad
Aut\'onoma de Zacatecas, Zacatecas Zac., MEXICO }
\address{$^3$Dual CP Institute of High Energy Physics, MEXICO}
\address{$^4$Departamento de F\'isica, Facultad de F\'\i sica 
e Inteligencia Artificial, Universidad Veracruzana, 91000 Xalapa, Veracruz, MEXICO}
\address{$^5$Departamento de F\'\i sica, CINVESTAV-I.P.N., Apdo. Postal 14740, 07000, 
M\'exico D.F., MEXICO}

\eads{
\mailto{cordero@esfm.ipn.mx},
\mailto{amolgado@fisica.uaz.edu.mx},
\mailto{efrojas@uv.mx}}

\begin{abstract}
 The early Dirac proposal to model the electron
 as a charged membrane is reviewed. 
A rigidity term, instead of the natural membrane tension, 
involving linearly the extrinsic
curvature of the worldvolume swept out by the membrane
is considered in the action modelling the bubble in the presence
of an electromagnetic field. We set up this model as a genuine
second-order derivative theory by considering a non-trivial 
boundary term which plays a relevant part in our formulation. 
The Lagrangian in question is linear in the bubble acceleration and by
means of the Ostrogradski-Hamiltonian approach we observed that the theory
comprises the management of both first- and second-class constraints. 
We show thus that our second-order approach is robust 
allowing for a proper quantization. We found an effective quantum
potential which permits to compute bounded states for the system. 
We comment on the possibility of describing 
brane world universes by invoking this kind of second-order correction terms.
\end{abstract}



\section{Introduction}

The pioneering Dirac proposal \cite{dirac} (and even before by
Less \cite{less}) to model the electron as a charged bubble,
started an exhaustive study of geometrical theories of
surfaces moving in a spacetime modelling some relevant physical
systems. This spinless Dirac geometrical theory describes a dynamic 
spherical membrane in the presence of an electromagnetic field 
where the non-electromagnetic forces are described by a constant 
surface tension. Unfortunately, as it was originally formulated, 
the model results limited in its predictive power.
Over the years the model has been improved by taking into
account the inclusion of second-order correction terms built 
from the extrinsic curvature of the worldvolume swept out by 
the bubble \cite{tucker1,tucker2}. Extrinsic curvature 
terms appear in several effective actions aimed to describe 
surfaces in diverse contexts which accommodate relativistic 
extended objects as notable realizations of interesting physical 
systems\footnote{As a matter of fact, this is the criterion for rigidity
that we follow in this work: It is a correction to Dirac-Nambu-Goto
objects in order to penalize singularities in their evolution thus
favoring a variety of configurations much rich in geometrical and
physical structure.} \cite{gregory,carter-greg}. In addition, it is well known that the 
inclusion of extrinsic curvature terms into geometrical models 
is reflected on the spin content of a given theory. In that sense, 
an effective model for the electron viewed as a charged spherically 
membrane and based on the inclusion of the extrinsic curvature 
of the worldvolume swept out by the bubble was developed by
\"Onder and Tucker (OT), in contrast to the original Dirac's 
model \cite{tucker1,tucker2}. Regrettably, the quantum approach 
performed for this model has the drawback of an unavoidable 
transcendental system which conceals the true dynamics of the 
system, and complicates its quantization. This model was also 
addressed in an harmonic estimate by studying radial modes~\cite{tucker2}.
The present status of this approach undoubtedly needs refinements 
since it only comprises a semi-classical approximation inherited 
from the transcendental equations arising for the aforementioned 
cumbersome phase space treatment. We are aware that this model 
does not pretend to be a realistic model for the electron, but, 
despite its plainness, it provides sufficient complexity in certain 
particle physics models, and also it introduces  interesting 
resemblances to cosmological brane models which deserves a careful analysis~\cite{guendelman,DR}.

In this paper we bring back to life the  model proposed in 
\cite{tucker1,tucker2}, retaining the basic physical idea of 
studying an elementary particle viewed as a $(2+1)$-dimensional 
gravitating electrically charged membrane in the presence of an 
electromagnetic field. Instead of the natural tension of the membrane,
to balance the Coulomb repulsion we consider a quite subtle surface 
stress derived from a correction term involving a linear extrinsic 
curvature term of the membrane. This type of terms has drawn
attention for a long time, specially in reference to the approach 
in the hypersurfaces context by Chen~\cite{yen}, and by Barrabes {\it et al}
as a correction term to study the dynamics of domain walls~\cite{barrabes}. 
Also, in order to achieve our goal, we consider an inspired alternative Hamiltonian 
approach by considering non-dynamic boundary terms. We overcome the 
usual quantization shortcomings for second-order derivative systems 
with the help of a canonical transformation and then we proceed to 
quantize this rigid membrane model. It is becoming more and more 
important to extended the powerful techniques of the Dirac treatment 
for constrained systems in describing physics with correctness, so that
we would like to improve the former quantum OT approach by means of 
a non-standard strategy for the classical canonical scheme relying on the Ostrogradski-Hamiltonian formalism. In this way, the appropriate 
classical phase space of the system is identified and worked out by 
the above mentioned canonical transformation which reveals the true 
nature of the system:  we are left with two first class and two 
second-class constraints, the former being of physical significance
while the latter are treated as identities by imposing the Dirac bracket.
At the quantum level, a Schr\"odinger-like equation is found and the 
effective potential is explicitly identified. Also, we have gained 
enough control of the global properties of this potential. Remarkably, 
we are able to show that the model under consideration leads to bounded states.

The paper is organized as follows.  Section~\ref{sec:model} 
introduces the geometrical model by considering an action governing 
the dynamics of a  membrane including a second-order correction term
via the extrinsic curvature.  In Section~\ref{sec:shell}, we specialize 
the model to an spherical shell, and discuss the nature of the 
Lagrangian components. Section~\ref{sec:ostro} and Section~\ref{sec:constraints}
are devoted to the analysis of the Ostrogradski-Hamiltonian
formalism for our system, and to the classification of the
emerging constraints.  In Section~\ref{sec:gauge} we go to the reduced 
phase space and perform a canonical transformation which reveals the 
true nature of the classical constraints.  We develop the quantum 
counterpart for our model in Section~\ref{sec:quantum}.  Finally, we 
present a summary and  some concluding remarks in Section~\ref{sec:conclusions}.

\section{The geometrical model}
\label{sec:model}

Consider a $2$-dimensional surface, $\Sigma$, evolving in a Minkowski 
$4$-dimensional background spacetime with metric $\eta_{\mu \nu}$, 
described by the embedding $x^{\mu}=X^{\mu}(\xi^{a})$ where $x^{\mu}$ 
are local coordinates for the background spacetime  ($\mu ,\nu= 0,1,2,3$), 
$\xi^{a}$ are local coordinates for the worldvolume, $m$, swept out 
by the surface ($a,b=0,1,2$) and $X^{\mu}$ are the embedding functions  
for $m$. We consider the following effective action underlying the 
dynamics of the surface $\Sigma$
\beq
S[X^\mu] = \int_m d^3\xi \,\left( 
- \alpha \sqrt{-g}\,K  +  \beta \,j^a \,e^\mu {}_a \,
A_\mu \right) \,,
\label{eq:action}
\eeq
where $K$ is the mean extrinsic curvature of the worldvolume
constructed with the extrinsic curvature tensor
$K_{ab}=-\eta_{\mu \nu}n^\mu D_a e^\nu{}_b $ and $g$ denotes the 
determinant of the induced worldvolume metric $g_{ab}= \eta_{\mu \nu}
e^{\mu}{}_a e^{\nu}{}_b$, where $e^\mu{}_a = X^\mu {}_{,\,a}$ 
are the tangent vectors to the worldvolume; $n^\mu$ is the 
spacelike unit normal vector to the worldvolume. Furthermore, $D_a = 
e^\mu{}_a D_\mu$, where $D_\mu$ is the background covariant 
derivative. The factor $\alpha$ is a constant related to the 
rigidity parameter of the surface $\Sigma$, and $\beta$ is the 
form factor of the model. Further, $A_\mu (x)$ is the gauge 
field living in the ambient spacetime, and $j^a$ is a fixed 
electric charge current density continuously distributed over 
the worldvolume, responsible for the minimal coupling between 
the charged surface and the electromagnetic field $A_\mu$ 
\cite{barut}. The action functional (\ref{eq:action}) is invariant under 
reparametrizations of the worldvolume $m$. 
Here, we regard $j^a$ as a function depending only on the worldvolume
coordinates  and, not being a true vector on the worldvolume it is 
locally conserved on $m$, $\partial_a j^a=0$. 
From a more generic perspective, the charge 
current density may be thought of a new dynamical variable, derivable from
a suitable internal potential living on $m$, as developed by Davidson
and Guendelman in~\cite{davidsonguendelman}.   We will not pursue this 
case in the rest of the paper.

The variation of the action functional with respect to the embedding 
functions~\cite{defo} leads us to the equation of motion
\beq
\alpha\,{\cal R} = \frac{\beta}{\sqrt{-g}}\,j^a\,n^\mu e^\nu {}_a 
 F_{\mu \nu}\,,
\label{eq:moto1}  
\eeq
which is a sort of  Lorentz-force law.  Here ${\cal R}$ is the worldvolume 
Gaussian curvature, and $F_{\mu \nu} = 2\partial_{[\mu} A_{\nu ]}$ is the 
strength tensor of the electromagnetic field. It is worthy notice that 
$j^a$ must remain unchanged under arbitrary deformations of 
the surface, $X \to X + \delta X$. Despite the 
effective Lagrangian density in equation~(\ref{eq:action}) is of second-order 
in derivatives, the resulting equations of motion are of second-order. 
This is a remarkable situation since then we will not have to struggle 
with ghost-like issues in the quantum formulation of the model.

\section{A moving electrically charged bubble}
\label{sec:shell}

We turn now to specialize the definitions of the previous section 
to the description of a spherical membrane $\Sigma$. From now on, 
we consider a background Minkowski spacetime described by
$ds^2 = - dt^2 + dr^2 + r^2 d\theta^2 + r^2 \sin^2 \theta d\phi^2$. 
We consider the membrane $\Sigma$ to be orientable and topologically
identical to $\Sigma = S^2 \times \mathbb{R}$, that is,  a bubble 
described by the following parametrization
\begin{equation}
x^\mu = X^\mu (\tau,\theta,\phi)= (t(\tau),r(\tau),
\theta,\phi)\,,
\end{equation}
so that the induced metric on the worldvolume is $ds^2 = g_{ab}d\xi^a 
d\xi^b = -N^2 d\tau^2 + r^2 d\theta^2 + r^2 \sin^2 \theta d\phi^2$ 
where $N= \sqrt{\dot{t}^2 - \dot{r}^2}$, for simplicity, and the 
dot stands for derivative with respect to the parameter $\tau$. It 
is worth mentioning that $N$ corresponds to the lapse function in 
the ADM Hamiltonian approach for branes~\cite{hambranes}. The normal 
vector to the worldvolume is implicitly defined by $g_{\mu \nu}n^\mu 
e^\nu{}_a = 0, $ and $g_{\mu \nu}n^\mu n^\nu = 1$, and explicitly reads
\begin{equation}
\nn
n^\mu = \frac{1}{N}
\left( \dot{r},\dot{t},0,0 \right)\,.
\end{equation}
Furthermore, for this parametrization we also have the unit 
timelike normal vector $\eta^a = \frac{1}{N} (1,0,0)$ such that 
$g_{ab} \epsilon^a {}_A \eta^b = 0$ and $g_{ab}\eta^a \eta^b = 
-1$ where $\left\lbrace \epsilon^a {}_A ,\eta^a \right\rbrace $  
is the $\Sigma$ basis viewed from $\Sigma$ into $m$ and $A,B=1,2$,
(see \cite{hambranes} for more details). The non-vanishing components 
of the extrinsic curvature tensor take the following form
\begin{equation}
\nn
 K^\tau{}_\tau = \frac{1}{N^3}\left( \ddot{r}\dot{t} - 
\dot{r}\ddot{t} \right) \,, \qquad K^\theta {}_\theta =
K^\phi {}_\phi = \frac{\dot{t}}{r\,N}\,.
\label{eq:Ks}
\end{equation}
These components are useful to compute the associated mean extrinsic 
curvature and the Gaussian curvature given by\footnote{The intrinsic 
curvature can be computed from the contracted Gauss-Codazzi 
integrability condition, ${\cal R}=K^2 - K_{ab}K^{ab}$.}
\begin{eqnarray}
 K &=& \frac{1}{N^3 r}\left[ r (\ddot{r} \dot{t} -
\dot{r}\ddot{t}) + 2\dot{t}N^2\right] \,,\\
{\cal R} &=& \frac{2\dot{t}}{N^4 r^2} \left[ 
2r (\ddot{r} \dot{t} -
\dot{r}\ddot{t}) + \dot{t} N^2 \right]\,. 
\end{eqnarray}

By invoking the electromagnetic potential on the spherical shell 
to take the specific form $A_\mu = \left( -\frac{q}{r}, 0,0,0\right)$ 
where $q$ is the total electric charge on the shell, and fixing the 
electric current as $j^\tau = q\,\sin \theta$, the equation of motion 
(\ref{eq:moto1}) reduces to
\begin{equation}
\label{eq:eom}
\frac{d}{d\tau}\left(  \frac{\dot{r}}{\dot{t}}\right) = - 
\frac{N^2}{2r\dot{t}^3}\left( \dot{t}^2 - \frac{N^2 \beta 
q^2}{2\alpha r^2} \right). 
\end{equation}

From the action (\ref{eq:action}) we see that the effective Lagrangian 
density specialized to the bubble reads 
\begin{equation}
{\cal L}= \sin \theta \left\lbrace - \frac{\alpha r}{ N^2}
\left[ r \left( \ddot{r} \dot{t} - \dot{r} \ddot{t} \right) 
+ 2 \dot{t} N^2  \right] \right\rbrace  + \frac{\beta}{\sin \theta} 
j^\tau \dot{t} A_0.
\label{eq:lagdensity}
\end{equation}
Thus, our effective model for the electron, in terms of an arbitrary 
parameter $\tau$, is reduced to
\begin{equation}
\nn
S = 4\pi \int d\tau \, L (r,\dot{r},\ddot{r},
\dot{t}, \ddot{t})\,,
\end{equation}
where the Lagrangian is given by
\begin{equation}
L = - \alpha \frac{r^2}{N^2} \left( \ddot{r} \dot{t} - \dot{r} 
\ddot{t}\right) - 2\alpha r \dot{t} - \beta   \frac{q^2 \dot{t}}{r}.
\label{eq:lagrangian}
\end{equation}
In addition to the velocities $\dot{t}$ and $\dot{r}$, this Lagrangian 
depends also on their corresponding accelerations $\ddot{t}$ and 
$\ddot{r}$.  So we are dealing with a genuine second order derivative 
theory. Note that (\ref{eq:lagrangian}) can be rewritten as 
$L = L_b + L_d$, where
\begin{eqnarray}
L_b&=& - \frac{d}{d\tau}\left[\alpha \,r^2 {\mbox{arctanh}} 
\left( \frac{\dot{r}}{\dot{t}}\right) \right]   \,,
\\
L_d &=&  2\alpha\, r \,\,\dot{r} \,\, {\mbox{arctanh}}
\left( \frac{\dot{r}}{\dot{t}}
\right) 
- 2 \alpha\,r \dot{t} -
\beta \frac{q^2 \dot{t}}{r}   \,,
\label{eq:lagdensity1}
\end{eqnarray}
are respectively a boundary term and a true dynamic term. As customary, 
the boundary term can be neglected without affecting the bubble 
evolution in time. This issue was addressed in \cite{tucker2} by 
taking into account a semi-classical harmonic oscillator approach due 
to the high degree of difficulty in the bubble evolution encountered.  
However, our treatment will rely on considering explicitly  both terms, 
the boundary and the dynamic, confronting us with a Lagrangian depending 
up to the accelerations, hence evoking an Ostrogradski-Hamiltonian 
formalism for the treatment of the model.

\section{Ostrogradski-Hamiltonian approach}
\label{sec:ostro}

The more general aspects for a Hamiltonian analysis
of an extended object have been presented in detail
in \cite{hambranes}. Here we will apply those results.

The highest conjugate momenta to the velocities 
$\left\lbrace \dot{t}, \dot{r}\right\rbrace $ are 
given by
\begin{eqnarray}
P_t &=& \frac{\partial L}{\partial \ddot{t}}=
 \alpha \frac{r^2 \dot{r}}{N^2}  \,,
\\
P_r &=& \frac{\partial L}{\partial \ddot{r}}=
- \alpha \frac{r^2 \dot{t}}{N^2}   \,,
\end{eqnarray}
such that the highest momentum spacetime vector can be 
rewritten as
\begin{equation}
P_\mu = - \alpha \frac{r^2}{N}\,n_\mu .
\label{eq:P}
\end{equation}
Note that the momentum $P_\mu$ is directed normal to the 
worldvolume. This is a general issue for this type of brane
models \cite{hambranes}.

The conjugate momenta to the position variables
$\left\lbrace t, r\right\rbrace $ are 
\begin{eqnarray}
p_t &=& \frac{\partial L}{\partial \dot{t}}
- \frac{d}{d\tau} \left(\frac{\partial L}{\partial
\ddot{t}} \right)  
= - \frac{2\alpha r\dot{t}^2}{N^2} - \beta \frac{q^2}{r}  
=:- \Omega\,, 
\label{eq:pt}
\\
p_r &=& \frac{\partial L}{\partial \dot{r}}
- \frac{d}{d\tau} \left(\frac{\partial L}{\partial
\ddot{r}} \right) = \frac{2\alpha r \dot{t}\dot{r}}{N^2}\,,
\end{eqnarray}
respectively. Important to note is the fact that both
momenta, $p_t$ and $p_r$, are from a totally different
nature. Indeed, while the momentum $p_t$ is not influenced
at all by the surface terms the momentum $p_r$ is obtained
by two contributions: $\textbf{p}_r$ coming from the ordinary dynamical 
theory ($L_d$) and $\mathfrak{p}_r$ coming from the boundary term 
($L_b$). In this way, we can denote the momentum $p_r$ as \cite{RT}
\begin{equation}
 \label{eq:ppr}
p_r := \textbf{p}_r + \mathfrak{p}_r\,,
\end{equation}
where
\begin{equation}
 \label{eq:prD}
\textbf{p}_r = \frac{ 2\alpha r\dot{t}\dot{r}}{N^2} + 
2\alpha r \,{\mbox{arctanh}} \left( 
\frac{\dot{r}}{\dot{t}}\right) ,
\end{equation}
and
\begin{equation}
 \label{eq:prB}
\mathfrak{p}_r = - 2\alpha r  \,{\mbox{arctanh}} \left( 
\frac{\dot{r}}{\dot{t}}\right) \,.
\end{equation}
It is crucial to recognize~(\ref{eq:prD}) as the canonical momentum
worked out in \cite{tucker2}, while~(\ref{eq:prB}) stands for the momentum conjugated
to the $r(\tau)$-variable when considering as the Lagrangian
only the surface term. 

In order to see the geometrical nature of 
the physical momentum it will be convenient to write the 
kinetic momentum, $\pi_\mu = p_\mu - \beta \,q\,A_\mu$, 
as follows
\begin{equation}
\nn
\pi_\mu = \frac{2\alpha r\dot{t}}{N^2}  \,\dot{X}_\mu \,.
\end{equation}
The constant quantity $\Omega$, previously introduced in 
relation~(\ref{eq:pt}), is nothing but the conserved energy 
as we will see below. We can rearrange the energy 
expression~(\ref{eq:pt}) in order to get a master evolution 
equation
\begin{equation}
N^2 + \dot{r}^2 = \gamma N^2 r^2 H^2\,,
\label{eq:master} 
\end{equation}
where $\gamma = \gamma(r)= \left( \Omega\,r - \beta q^2\right)/r^4$,
and we have introduced $H^2=1/2\alpha$. This relation represents 
for the bubble model an analogous master equation emerging in the 
context of quantum geodetic brane gravity \cite{RT,R-T,Geodetic,davidson}.

Hitherto, the appropriate phase space of the system, $\Gamma = 
\left\lbrace t,r,\dot{t},\dot{r};p_t,p_r,P_t,P_r \right\rbrace$,  
has been explicitly identified. In order to complete the 
Ostrogradski-Hamiltonian programme in phase space $\Gamma$, we 
will consider the canonical Hamiltonian
\begin{eqnarray}
H_0 &=& p_r\,\dot{r} + p_t\,\dot{t} + P_r\,\ddot{r}
+ P_t\,\ddot{t} - L 
\nonumber \\
&=&\pi_r\,\dot{r} + \pi_t\,\dot{t} + J_{K}\,,
\label{eq:ham}
\end{eqnarray}
where we have defined
\begin{equation}
J_{K} := \sqrt{2\alpha} N \left( \Omega \,r - \beta \,q^2\right)^{1/2} \,.
\end{equation}
As expected, the canonical Hamiltonian results a function only 
of the physical momentum $\pi_\mu$. It may look like an unnecessary 
complication to write both the physical momentum and $H_0$ in terms 
of $\Omega$, but this quantity results important for the reason 
that it is merely the conserved internal energy.  In order to 
demonstrate this, we only have to consider the Poisson bracket 
(defined below in relation~(\ref{eq:PB})) of the momentum $p_t$ with 
the canonical Hamiltonian $H_0$.  

\subsection{Classical equilibrium configuration}
\label{sec:ostro1}

We now briefly describe the equilibrium conditions of the system. 
The radius of the static solution, $r_0$, is obtained if we put 
$\dot{r} = \ddot{r}=0$ in Eqs. (\ref{eq:eom}) and (\ref{eq:pt}). 
Thus, we have
\numparts
\begin{eqnarray}
0 &= 2\alpha - \frac{\beta q^2}{r_0 ^2},
\label{eq:condi-1}
\\
\Omega & = 2\alpha r_0  + \frac{\beta q^2}{r_0},
\label{eq:condi-2}
\end{eqnarray}
\endnumparts
where (\ref{eq:condi-2}) is the static energy in equilibrium. It is 
worth mentioning that these conditions are gauge independent. Further, 
note that $d\Omega/d r_0 = 0$ implies Eq. (\ref{eq:condi-1}), which is 
a consequence of the minimum for the classical potential 
function. The condition (\ref{eq:condi-2}) specialized to the electron 
properties and $\Omega=m_e$, lead to the result\footnote{We will work in natural 
units, $\hbar = c =1$.} 
\beq
r_0 = \frac{\lambda}{m_e} \qquad \mbox{and} \qquad \alpha = \frac{\lambda}{4r_0 ^2},
\eeq
where $\lambda:= e^2 =1/137$ is the structure constant, $e^2:=q^2$ 
and $\beta = 1/2$.

\section{First- and second-class constraints}
\label{sec:constraints}

As already mentioned, we are dealing with a second-order derivative theory
because, in addition to the velocities, the Lagrangian (\ref{eq:lagrangian})
depends also on the accelerations. From the definition of the 
momenta (\ref{eq:P}) we can get the primary constraints
\begin{eqnarray}
{\cal C}_1 &=& P \cdot \dot{X} \approx 0 \,,
\label{eq:C1}
\\
{\cal C}_2 &=& N P\cdot n + \alpha r^2 \approx 0\,,
\label{eq:C2}
\end{eqnarray}
where the dot symbol denotes contraction with the Minkowski 
metric.\footnote{Note that, in general, $n^\mu$  is a function 
of the derivatives with respect to 
the parameter $\tau$ of the embedding functions $X$.} 
Also, the $\approx$ symbol stands for weak equality in the Dirac approach for
constrained systems \cite{dirac1,teitelboim}.  The previous
constraints are supported by the completeness 
relation 
\begin{equation}
\nn
\eta^{\mu \nu} = n^\mu n^\nu - \eta^\mu \eta^\nu 
+ h^{AB} \epsilon^\mu {}_A \epsilon^\nu {}_B\,,
\label{eq:Hmunu}
\end{equation}
satisfied by the $\Sigma$ basis, $ \left\lbrace 
\epsilon^\mu {}_A, \eta^\mu, n^\mu \right\rbrace $,
where $h_{AB} = \eta_{\mu \nu} \epsilon^\mu {}_A \epsilon^\nu {}_B$ is the spatial
metric on $\Sigma$ and $\epsilon^\mu {}_A$ 
the corresponding tangent vectors to $\Sigma$. 
The vector $\eta^\mu$ denotes a timelike unit normal 
vector to $\Sigma$~\cite{hambranes}.

The total Hamiltonian that generates time evolution
is given by
\begin{equation}
H_T = H_0 + \lambda_1 {\cal C}_1 + \lambda_2 {\cal C}_2   \,,
\end{equation}
where $\lambda_1$ and $\lambda_2$ are Lagrange multipliers
enforcing the primary constraints ${\cal C}_1$ and ${\cal C}_2$.

Time evolution for any canonical variable $z \in \Gamma$ 
is thus dictated by means of
\begin{equation}
\dot{z} = \left\lbrace z, H_T \right\rbrace\,, 
\end{equation}
where the corresponding Poisson brackets (PB) for any two
functions $F(z), G(z) \in \Gamma$ is appropriately
defined as
\begin{eqnarray}
\left\lbrace F,G \right\rbrace &:=& \frac{\partial F}{\partial
t}\frac{\partial G}{\partial p_t} +  \frac{\partial F}{\partial
a}\frac{\partial G}{\partial p_a} +
\frac{\partial F}{\partial \dot{t}}\frac{\partial G}{\partial P_t}
+ \frac{\partial F}{\partial \dot{a}}\frac{\partial G}{\partial P_a}
- (F \leftrightarrow G)  \,.
\label{eq:PB} 
\end{eqnarray}
Note that $\left\lbrace {\cal C}_1,{\cal C}_2 \right\rbrace 
= 0$ and in consequence the primary constraints (\ref{eq:C1})
and (\ref{eq:C2}) are in involution.
As primary constraints must be 
preserved in time, that is, $\dot{{\cal C}}_1\approx 0$ 
and $\dot{{\cal C}}_2\approx 0$, we are lead to the secondary constraints
\begin{eqnarray}
{\cal C}_3 &=& H_0 \approx 0 \,,
\label{eq:C3}
\\
{\cal C}_4 &=& N\pi \cdot n \approx 0 \,.
\label{eq:C4}
\end{eqnarray}
The vanishing of the canonical Hamiltonian is expected as for any 
reparametrization invariant theory. Geometrically, 
the canonical Hamiltonian $H_0$ generates diffeomorphisms normal 
to the worldvolume. The secondary constraint (\ref{eq:C4}) is 
characteristic of every brane model linear in accelerations 
\cite{RT}. The process of generation of further constraints is 
stopped at this stage since ${\cal C}_3$ is automatically 
preserved under time evolution 
and the requirement of being stationary for ${\cal C}_4$ only determines 
a restriction on one of the Lagrange multipliers, namely, $\lambda_2 
= - \left( \dot{t} ^2 -  \frac{\beta q^2 N^2 }{2\alpha r^2} 
\right)/2 r \dot{t}$. Thus, we are dealing with an entirely 
constrained theory with two primary and two secondary constraints.

Following Dirac's programme, the set of constraints should be 
separated into subsets of first- and second-class constraints
\cite{dirac1,teitelboim} (see also \cite{RT,Geodetic} in the context of geodetic brane 
gravity). 
For our system we have two first-class constraints
and two second-class constraints.  We judiciously choose them as 
\begin{eqnarray}
{\cal F}_1 & := &  {\cal C}_1 \,, 
\label{eq:f1} 
\\
{\cal F}_2 & := & - \frac{\left( \dot{t} ^2 -  \frac{\beta q^2 N^2 }{2\alpha r^2} 
\right) }{2 r \dot{t} }
 {\cal C}_2 + {\cal C}_3\,,
\label{eq:f2}
\\
{\cal S}_1 & := & {\cal C}_2 \,, 
\label{eq:s1} \\
{\cal S}_2 & := & {\cal C}_4 \,,
\label{eq:s2}
\end{eqnarray}
where the ${\cal F}$'s and the ${\cal S}$'s stand respectively as 
the first- and the second-class constraints. 
In order to eliminate the extra degrees of freedom in our canonical approach
we must replace the PB with the Dirac brackets (DB) defined by 
\beq
\left\lbrace F,G \right\rbrace^{*}:= \left\lbrace  F,G\right\rbrace - \left\lbrace F, {\cal S}_i
\right\rbrace {\cal S}_{ij} ^{-1} \left\lbrace {\cal S}_j , G \right\rbrace \,,
\eeq
where ${\cal S}_{ij} ^{-1}$ stands for the inverse elements
of the second-class constraints matrix defined by ${\cal S}_{ij}:= \left\lbrace {\cal S}_i ,{\cal S}_j 
\right\rbrace $ , \quad ($i,j= 1,2.$). We find straightforwardly in our case the matrix
\beq
\left( {\cal S}_{ij} \right) \approx 4\alpha r \dot{t} \left( 
\begin{array}{cc}
0 & 1
\\
-1 & 0
\end{array}
\right) .
\eeq
We may therefore construct the DB, and we find
\beq
\left\lbrace F,G \right\rbrace^{*}= \left\lbrace  F,G\right\rbrace +
\frac{1}{4\alpha r \dot{t}} \left( \left\lbrace F , {\cal S}_1 \right\rbrace 
\left\lbrace  {\cal S}_2 , G \right\rbrace 
- \left\lbrace F , {\cal S}_2 \right\rbrace 
\left\lbrace  {\cal S}_1 , G \right\rbrace
\right) 
\eeq
Hence, we consider the second-class constraints to vanish strongly which helps
to eliminate the part proportional to ${\cal C}_2$ in (\ref{eq:f2}) leading thus to a
more suitable expression form for ${\cal F}_2$. The final first-class 
Hamiltonian for our model is $H = \alpha^A {\cal F}_A$, \quad $(A,B=1,2)$.
We turn now to the counting of physical degrees of freedom (dof).
According to the recipe developed in \cite{teitelboim}, the model has 
$(8 - 2 \times 2 - 2)/2 = 1$ dof.
Note that as we have two linear independent first-class constraints, we will have the 
presence of two gauge transformations for this brane model. We will  
analyze the gauge-fixing in the next section.

\section{Gauge--fixing}
\label{sec:gauge}

According to the conventional Dirac scheme, in order to extract the physical 
meaningful phase space for a constrained system we need a gauge--fixing prescription 
which entails the introduction of extra constraints, avoiding in this way the 
gauge freedom generated by first-class constraints (\ref{eq:f1}) and (\ref{eq:f2}). 
To achieve
this we will consider the following gauge condition
\begin{equation}
\label{eq:varphi1}
 \varphi_1 = N - 1= \sqrt{\dot{t}^2 - \dot{r}^2} - 1 \approx 0\,,
\end{equation}
and the generalized evolution equation (\ref{eq:master})
\begin{equation}
\label{eq:varphi2}
\varphi_2 =  N^2 + \dot{r}^2 - \gamma\,N^2 H^2 r^2 \approx 0\,.
\end{equation}
From the geometric point of view, this set of gauge conditions is good enough 
since the matrix $\left(\left\{ {\cal F} , \varphi_{1,2} \right\}\right)$ is 
non-degenerate in the constraint surface. Indeed, under the Poisson bracket 
structure~(\ref{eq:PB}), it is straightforward to show that gauges $\varphi_1$ 
and $\varphi_2$ form a second-class algebra with the constraints ${\cal F}_1$ 
and ${\cal F}_2$
\begin{eqnarray}
\nn
\left\lbrace \varphi_1, {\cal F}_1 \right\rbrace &=& \varphi_1 + 1  \,, 
\\
\nn
 \left\lbrace \varphi_1, {\cal F}_2 \right\rbrace &=&0   \,, 
 \\
 \nn
 \left\lbrace \varphi_2, {\cal F}_1 \right\rbrace &=& 2\varphi_2 \,, 
\\
\nn
 \left\lbrace \varphi_2, {\cal F}_2 \right\rbrace &=&  - 
\frac{\left( \dot{t} ^2 -  \frac{\beta q^2 N^2 }{2\alpha r^2} 
\right)}{r}
 \dot{r}  \,.
\end{eqnarray}
Consequently, velocities $\dot{t}$ and $\dot{r}$ must be discard as dynamical degrees 
of freedom.

Next, in order to get control over the model, we implement 
the following canonical transformation to a new set of phase space 
variables
\begin{eqnarray}
N     & :=&  \sqrt{\dot{t}^2 - \dot{r}^2}\,,   \\
\Pi_N & :=&  \frac{1}{N}(P\cdot \dot{X}) \,,   \\
\nu     & :=& - \left[  N (P\cdot n) + \alpha\,r^2 \right] , \\
\Pi_\nu & :=&  \mathrm{arctanh} \left( \frac{\dot{r}}{\dot{t}}\right) \,,
\label{eq:transf}
\end{eqnarray}
together with the transformation 
\begin{eqnarray}
X^\mu &=& X^\mu\,,
\nonumber
\\
\textbf{p}_\mu &:=& p_\mu + \left\lbrace p_\mu , \nu \right\rbrace
\Pi_\nu \,.
\label{eq:phys-p}
\end{eqnarray}
Important to note is that in this canonical transformation the coordinates 
$X^\mu$ remain unaltered, while the dynamical momentum 
$\textbf{p}_r$ is distinguished as the relevant momentum of the model.
Such transformation can be physically interpreted as a Lorentz rotation in 
phase space which, straightforwardly, preserves the structure of the canonical 
Poisson brackets
\beq
 \left\lbrace N , \Pi_N \right\rbrace &=&  1= \left\lbrace \nu , \Pi_\nu \right\rbrace, 
 \\
 \left\lbrace X^\mu  , \textbf{p}_\nu \right\rbrace &=& \delta^\mu {}_\nu\,,
\eeq
as expected.

In terms of the new phase space variables, the first- and 
second-class constraints (\ref{eq:f1})-(\ref{eq:s2}) become
\begin{eqnarray}
\label{eq:F1}
\fl
{\cal F}_1 &=& N \Pi_N = 0\,, \\
\label{eq:F2}
\fl
{\cal F}_2 &= & N \left[ \left( p_t + \beta \frac{q^2}{r} \right)\cosh \Pi_\nu
+ \left( \textbf{p}_r - 2\alpha r \Pi_\nu\right)\sinh \Pi_\nu 
+ 2\alpha r \cosh \Pi_\nu \right] = 0 \,, 
\end{eqnarray}
and
\begin{eqnarray}
 {\cal S}_1 &=& \nu =0  \,,
\label{eq:SS1} \\
{\cal S}_2 &=& N \left[ \left( p_t + \beta \frac{q^2}{r} \right)\sinh \Pi_\nu
+ \left( \textbf{p}_r - 2\alpha r \Pi_\nu\right)\cosh \Pi_\nu 
 \right] = 0    \,,
\label{eq:SS2}
\end{eqnarray}
respectively.  Note that in the second-class constraint~(\ref{eq:SS2}) 
we split the momentum conjugated to $r$ according to relations~(\ref{eq:ppr})
and (\ref{eq:phys-p}). As customary, 
the second-class constraints (\ref{eq:SS1}) and (\ref{eq:SS2}) 
may be taken as algebraic identities after implementing the 
Dirac bracket~\cite{dirac1,teitelboim}.
Furthermore, these second-class identities will become auspicious at 
the quantum level since they enclose important operator identities. 

The constraint ${\cal F}_1$ is simply associated to the 
gauge transformations $N\partial_N-\Pi_N \partial_{\Pi_N}$ which 
only acts on the $N\Pi_N$-plane.  
As for  the constraint  ${\cal F}_2$ in equation~(\ref{eq:F2}), we can further transform it by 
expressing the hyperbolic functions in terms of the phase space variables as follows, 
\beq
\label{eq:cosh}
\cosh \Pi_\nu &=& - \frac{\left( p_t + \beta \frac{q^2}{r}\right)}{r^2 (2\alpha \gamma)^{1/2}}  \,,\\
\label{eq:sinh}
\sinh \Pi_\nu &=& \frac{\left( \textbf{p}_r - 2\alpha r 
\Pi_\nu\right) }{ r^2 (2\alpha \gamma)^{1/2}}\,,
\eeq 
where we have substituted the 
second gauge condition~(\ref{eq:varphi2}) into the second-class constraint (\ref{eq:SS2}). 
Thus, ${\cal F}_2$ is transformed into
\beq 
\label{eq:finalF2}
\hspace{-12ex}
{\cal F}_2 &=& \frac{N}{ r^2 (2\alpha\gamma)^{1/2}} \left[ \left( \textbf{p}_r -
2\alpha r \Pi_\nu \right)^2 - \left( p_t + \frac{\beta q^2}{r} \right)^2 
 - 2 \alpha r  \left( p_t + \frac{\beta q^2}{r} \right) \right] =0  \,,
\eeq
and we have arrived to an expression  quadratic in 
the physical momenta for the canonical
Hamiltonian $H_0$, which is identified with the constraint ${\cal F}_2$ 
when the second-class constraint ${\cal S}_1$ is considered.  

In order to remark the relevance of the new canonical variables,
we observe that the internal energy $\Omega$ given by 
expression~(\ref{eq:pt}) can be written as $\Omega=
2\alpha r \cosh^2 \Pi_\nu+\beta q^2/r$ which at first-order approximation
matches the potential energy in~\cite{tucker2} in the so-called low velocity 
regime.  However, it is easy to show that an additional kinetic term 
$\frac{\Pi^2 _\nu}{2\mu(r)} $ emerges
from this expression by considering up to second-order in the expansion of 
$\cosh\Pi_\nu$, where $\mu(r)=(4\alpha r)^{-1}$.  Note that this mass-like 
term is a function of $r$. In addition, if we consider the first-class 
constraint ${\cal F}_2$, we can obtain the expression $\Omega=\pm\sqrt{(\textbf{p}_r -
2\alpha r \Pi_\nu)^2+m^2(r)}+V_{\mathrm{rel}}(r)$, where the mass-like term is given 
by $m^2(r):= \alpha^2 r^2$, and the 
relativistic potential 
is given by $V_{\mathrm rel}(r):=\beta q^2/r + \alpha r$.

\section{Quantum approach}
\label{sec:quantum}

In this section we study the canonical quantization of our system.  
To this end, we emphasize the totally 
dissimilar nature which first- and second-class constraints play in the quantum 
theory.  We start with the conventional way by promoting 
the classical constraints into operators.
However, by implementation of the Dirac bracket in the classical counterpart 
we are able to eliminate second-class constraints off the theory  by converting 
them into strong identities.  
At the quantum level this is mirrored by defining the 
quantum commutator of two quantum operators, $\hat{A}$ and $\hat{B}$, as 
\beq
\nn
\label{eq:commutator}
[\hat{A},\hat{B}]:=i\widehat{\lbrace A,B \rbrace^\ast} \,,
\eeq
where $\lbrace\cdot,\cdot\rbrace^\ast$ stands for the Dirac 
bracket. 
Thus, with this prescription the operators 
corresponding to second-class constraints are also enforced as operator 
identities~\cite{dirac1,teitelboim}.  For the system in question, this yields the quantum operator 
expressions
\beq
\hat{{\cal S}}_1 &:=& \hat{\nu} =0   \,,
\label{eq:SSq1} \\
\hat{{\cal S}}_2 &:=& \hat{N} \left[ \widehat{\left( p_t + \beta \frac{q^2}{r} \right)} 
\widehat{\sinh \Pi_\nu}
+ \widehat{\left( \textbf{p}_r - 2\alpha r \Pi_\nu\right)}
\widehat{\cosh \Pi_\nu} 
 \right] = 0 \,,
\label{eq:SSq2}
\eeq
which, in particular, tell us the character of the quantum operators
$\hat{\nu}$, $\widehat{\cosh \Pi_\nu}
$ and $\widehat{\sinh \Pi_\nu}$ (see equations (\ref{eq:cosh}) and (\ref{eq:sinh})). 
Also, we will represent the radial operator  as $\hat{\textbf{p}}_r:=-i(\partial/\partial r)r$ since 
then the operator $\hat{\textbf{p}}^2_r=-(\partial^2/\partial r^2 +(2/r)\partial/\partial r)$ will be 
Hermitian in the inner product of states in a conventional Hilbert
space, namely an $L^2$-space. 
For the rest of the variables, we choose 
to work on the ``position'' representation, where we consider the position 
operators by multiplication and their associated momenta operators by $-i$ times 
the corresponding derivative operator.

\begin{figure}
\begin{center}
\includegraphics[angle=0,width=14.2cm,height=9.5cm]{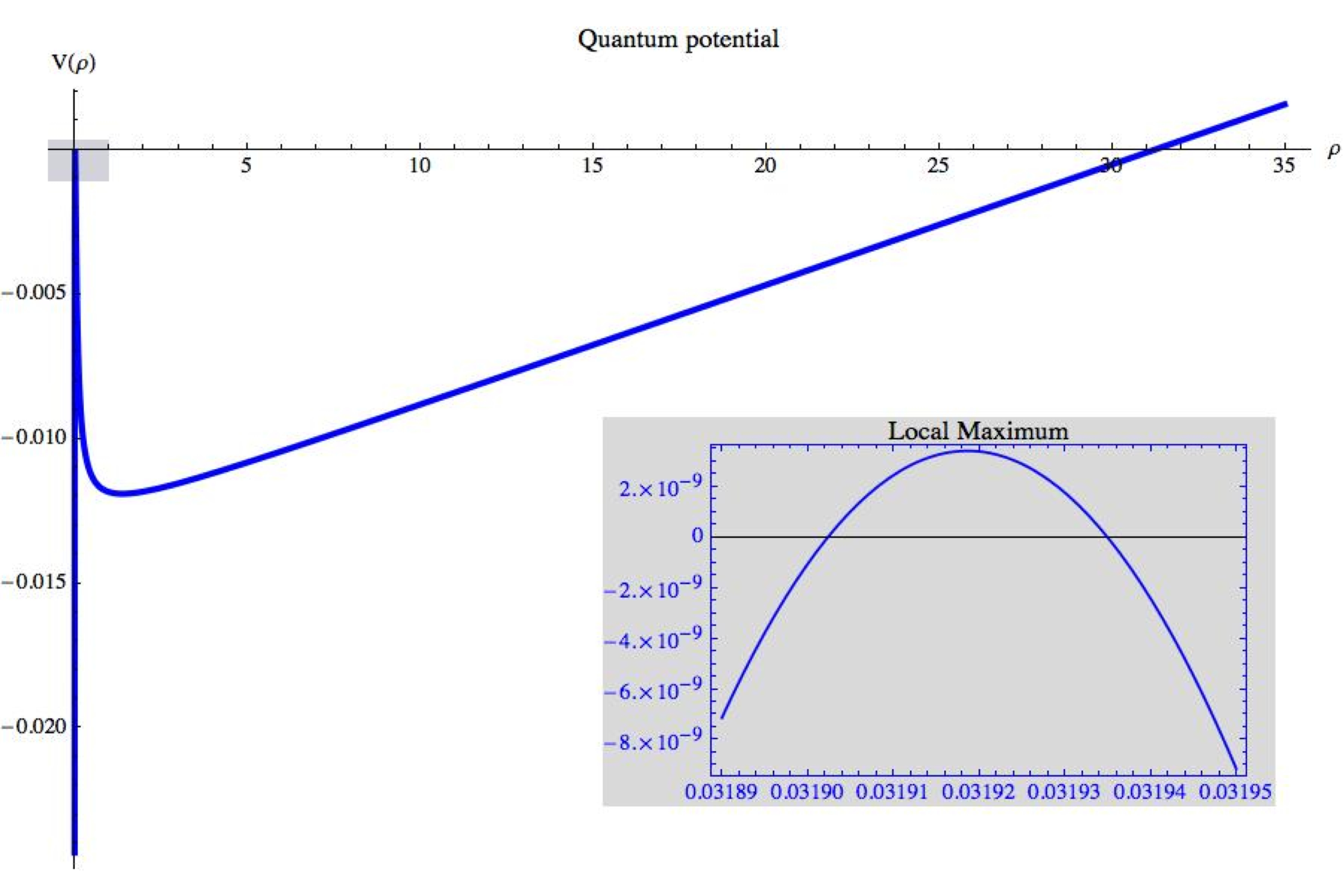}
 \caption[short caption for figure 11]{\label{fig:potential} {Effective quantum 
potential of a charged rigid bubble for large values of $\varrho$ and when $\varrho$ approaches to zero. For this graph we considered the energy value $\varepsilon =
0.1144817$.}}
\end{center}
\end{figure}

Next, we will adopt as our quantum first-class constraints the operators
\beq
\label{eq:quantumF1}
\hat{{\cal F}}_1 &:=& -iN\frac{\partial\ }{\partial N}    \,,\\
\label{eq:quantumF2}
\hat{{\cal F}}_2 &:=& N \left[ \left( \hat{\textbf{p}}_r -
2\alpha r \hat{\Pi}_\nu \right)^2 - \left( \hat{p}_t + \frac{\beta q^2}{r} \right)^2 
 -2\alpha r \left(\hat{p}_t + \frac{\beta q^2}{r} \right) \right]  \,.
\eeq
Note that the $N$ factor in $\hat{\cal{F}}_2$ is necessary in order to 
maintain at the quantum level the classical algebraic structure between the  two first-class 
constraints.   Also note, that for simplicity, we have chosen a trivial factor ordering 
which allowed us to get rid of  the denominator in 
equation~(\ref{eq:finalF2}).\footnote{Indeed, if we consider a different factor 
ordering for the quantum counterparts of first-class 
constraints, (\ref{eq:F1}) and (\ref{eq:finalF2}), we have as a result that the 
wave function is a homogeneous function of degree minus one half in the $N$-variable, 
and also, the potential depending on the $r$-variable (described below in 
(\ref{eq:potential})) includes an extra inverse square term.  Important to mention is the fact that 
the common denominator in~(\ref{eq:finalF2}) is cancelled out in the low 
velocity regime of~\cite{tucker2}, and hence the factor ordering they have chosen 
results simpler in that case.  From a more extensive point of view, the 
operator ordering problem can be tackled by following the arguments developed in quantum cosmology
as done in~\cite{Guenkaga} and~\cite{checos}, where the operator ordering is related to the 
avoidance of singularities and to the Hermiticity of the Hamiltonian operator. In fact, among
the most popular choices (not unique) for a suitable factor ordering in quantum cosmology, 
is to consider expressions with derivatives in the form of a Laplace-Beltrami operator.
At a fundamental level in our quantum description there is not obvious election for a factor
ordering. Our choice for the operator $\hat{\textbf{p}}_r$ results Hermitian and in 
consequence the implicit operator ordering ambiguities are fixed in the expression~(\ref{eq:quantumF2})
(for example, compare with Eq.~(A.1) in~\cite{Guenkaga} with $p=2$ in flat minisuperspace).}
 First, we explore the quantum  
equations emerging by considering the physical states $\Psi$ 
of the theory as those defined by na\"ive Dirac conditions
\beq
\label{eq:Dirac1}
\hat{{\cal F}}_1 \Psi & = &   0   \,,\\  
\label{eq:Dirac2}
\hat{{\cal F}}_2 \Psi & = &   0   \,.
\eeq
Equation~(\ref{eq:Dirac1}) simply tells us that our physical states $\Psi$ 
are not explicitly depending on the phase space variable $N$.
As classically, equation~(\ref{eq:Dirac2}) is the most interesting for us, since it is 
related to the Hamiltonian operator $\hat{H}_0$, hence resulting in a Schr\"odinger-like 
equation. 
In order to find solutions to this equation, we notice first that the second-class constraint $\hat{\mathcal{S}}_1$
tell us that the variable $\nu$ is fixed to zero, thus getting rid of any possible 
dependence on this variable in $\Psi$, which in turn leads to the conclusion 
that the action of the operator $\hat{\Pi}_\nu$ on the $\Psi$ states vanishes automatically.
Further, we see that the $t$-dependence can be solved
by assuming $\Psi(r,t):=e^{-i\Omega t}\psi(r)$, in agreement with the classical 
definition for $\Omega$.  Finally, we notice that $\psi(r)$ must 
satisfy the differential equation
\beq
\label{eq:Schro}
\left[  -\left( \frac{d^2 }{d r^2} 
+ \frac{2}{r}\frac{d}{d r} \right)
+ V(r) 
 \right] \psi(r) = 0\,,
\eeq 
where the quantum potential $V(r)$ is given by
\begin{equation}
V (r) =  - \frac{\beta^2 q^4}{r^2} + \frac{2\beta \Omega q^2}{r}
+ 2\alpha \Omega \,r - \Omega^2 - 2\alpha \beta q^2 \,,
\label{eq:potential} 
\end{equation} 
The shape of this potential is depicted in Figure~\ref{fig:potential}. 
The real zeroes of $V(r)$ are located at
$r_1:= \frac{\beta q^2}{\Omega}$, $r_2:= \frac{\Omega}{4\alpha}\left( 1 - 
\sqrt{1 - \frac{8\alpha \beta q^2}{\Omega^2}} \right)$ and
 $r_3:= \frac{\Omega}{4\alpha}\left( 1 + 
\sqrt{1 - \frac{8\alpha \beta q^2}{\Omega^2}} \right)$ if the condition 
$\Omega^2 >8\alpha \beta q^2$ holds. We will work under this condition. 
It follows from the preceding expressions for the zeroes of the potential
that if $\Omega$ increases, $r_1$ and $r_2$ come to zero and, in 
consequence, the local maximum decreases. In opposition, as $\Omega$
increases the zero located at $r_3$  grows.
The potential tends to infinity as $r$ grows  whereas it becomes singular 
at minus infinity as $r$ is approaching to zero.
Note that this effective potential has a linearly raising dependence 
of the bubble radius which is different in comparison with the original 
Dirac model which possesses a quartic dependence in the bubble radius 
added with the first, the second and the fourth term of the potential 
(\ref{eq:potential}). The first term in the potential (\ref{eq:potential}) 
corresponds to a centrifugal-like potential whereas the second term is 
related to the usual Coulomb interaction and, the latter term may be 
conceived as a shift in the bubble energy levels by the quantity $2\alpha 
\beta q^2$. 

\noindent
The issue of computing explicit solutions of the Eq. (\ref{eq:Schro}) results
very complicated, instead we bring into a play a numerical technique based 
in a Fortran code in order to compute the energy spectrum. For this reason, 
it is convenient to choose the dimensionless quantities, $\varrho :=r/r_0$ 
and $\varepsilon = r_0 \Omega$ (see for example~\cite{Stedile}) where we have 
considered the results of the Subsection \ref{sec:ostro1}. Now, by a suitable 
alternative function $\phi(r) :=\psi(r)/r$, the radial equation (\ref{eq:Schro}) 
becomes
\beq
\left( -  \frac{d^2}{d\varrho^2} 
+ V(\varrho) \right)\phi(\varrho) = 0\,, 
\label{eq:chodinger}
\eeq
where
\beq
V(\varrho) = - \frac{\lambda^2}{4\varrho^2} + \frac{\varepsilon \lambda}{\varrho}
+ \frac{\varepsilon\lambda}{2} \,\varrho - \left( \varepsilon^2 +
\frac{\lambda^2}{4}\right)  \,.
\label{eq:Potential} 
\eeq

\begin{figure}
\begin{center}
\includegraphics[angle=0,width=14.2cm,height=9.5cm]{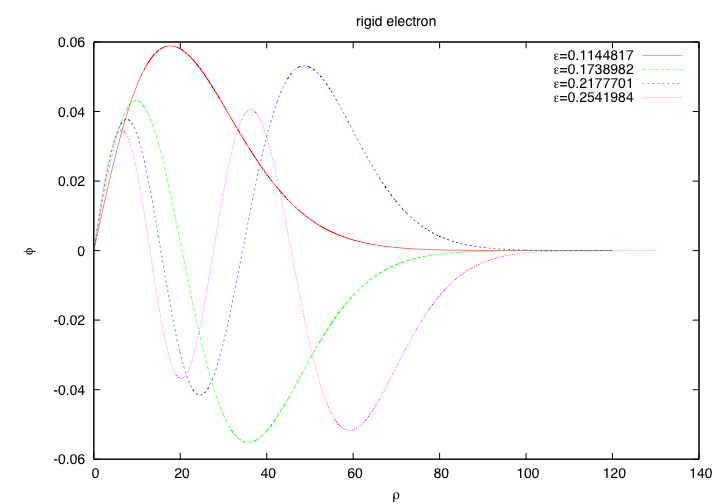}
 \caption[short caption for figure 3]{\label{fig:wavescolor} {The first normalized eigenfunctions of a charged rigid membrane.}}
\end{center}
\end{figure}

\noindent
By employing the fourth-order Runge-Kutta method altogether with the necessary
boundary conditions $\phi(0)=0$ and $\phi'(0)=A$, the latter being an arbitrary value which
is not relevant for the final result, 
we are able to obtain the solutions of the Eq. (\ref{eq:chodinger}), \cite{giordano}.
Now, by using the shooting method, both the eigenfunctions and
eigenvalues of the  Schr\"odinger-like equation (\ref{eq:chodinger}) can be computed. Some of the  
excited states eigenvalues of the charged bubble are shown in Table
\ref{teibel}. Similarly, in Figure \ref{fig:wavescolor} we plot the first normalized
eigenfunctions $\phi_n (\varrho)$. At this point, we are able to compare our results with
those developed by \"Onder and Tucker in~\cite{tucker2}. We found that for $n=1$, $\Omega_1 = 15.68 \,\,m_e$ and for $n=133$, 
$\Omega_{133} = 206.768\,\, m_e$ (the muon 
mass), unlike the values obtained by OT where, for instance, they found
for the ground state $n=1$, $\Omega_1\simeq 13\,\,m_e$ and for $n=190$, $\Omega_{190}
\simeq 200\,\,m_e$. This fact is a merely consequence of the several aspects
of the quantum approach developed by OT where a semiclassical quantization, in
a specific gauge, was adopted. Hence, the old idea pursued by Dirac regarding that 
the first excited state could be considered as a muon is not realized for this 
rigid bubble model.

\begin{table}
\caption{\label{label}Eigenvalues of the Schr\"odinger-like equation (\ref{eq:chodinger}) and the excited energies of the charged bubble}
\begin{indented}
\item[]\begin{tabular}{@{}ccc}
\br
$n$ & $\varepsilon_n$ & $\Omega_n /m_e$ \\
\mr
1 & 0.1144817 & 15.6839929 \\
2 & 0.1738982 & 23.8240616 \\
3 & 0.2177701 & 29.8345146 \\
4 & 0.2541984 & 34.8251903 \\
5 & 0.2860303 & 39.1861634 \\
6 & 0.3146605 & 43.1085008 \\
7 & 0.3408957 & 46.7027122 \\
8 & 0.3652517 & 50.0394897 \\
9 & 0.3880928 & 53.1686423 \\
10& 0.4096428 & 56.1210636 \\
\br
\end{tabular}
\end{indented}
\label{teibel}
\end{table}

\section{Concluding remarks}
\label{sec:conclusions}

In this paper we reviewed the classical and quantum aspects of 
a rigid charged conducting membrane.  We considered our model as 
a dynamical bubble in the presence of an electromagnetic
field with the non-electromagnetic forces described by a term 
depending linearly on the extrinsic curvature
of the worldvolume. As discussed before, 
though such correction term involves second-order derivatives, the
ensuing equations of motion are of second-order in the field variables,
as expected. We would like to remark that a generic minimal 
coupling term between the charge and the field $A_\mu$ has been 
considered in this work \cite{barut}, and it is equivalent to the Dirac original proposal 
that introduces a coupling by means of a boundary condition which 
is consistent only in a particular gauge \cite{dirac}.

Our analysis took into consideration the routinely neglected boundary 
term which involves second-order derivatives of the configuration variables.
Thus, we evoked the Ostrogradski-Hamiltonian formalism in order to 
investigate the constrained structure of the model.  Indeed,    
a canonical transformation
allowed us to acquire control over our set of constraints..
As expected, we are ushered to a Hamiltonian constraint which do not depend on 
the momenta associated to the higher-order variables, hence the 
structures coming from the boundary term do not play a relevant 
part on the dynamics of the system, but nonetheless they play the important 
role of being a bridge to obtain quadratic expressions in the physical 
momenta, useful for a passage to the quantum theory.

At the quantum level, we considered the canonical formalism of 
Dirac in order to overcome certain issues found in previous attempts by 
several other authors.  In particular, we obtained a Schr\"odinger-like
equation where the explicit effective quantum potential is identified, and the 
global behavior of the potential is discussed. As a byproduct, we have made a
reappraisal of the excited states eigenvalues of this type of charged membrane
which demonstrates the significant value of prediction of the scheme followed here.

More comments are in order. The Dirac model presents signals of instability 
 \cite{hungaros}. Perhaps, this shortcoming could be overcome if we 
take into account a linear extrinsic curvature correction term in the model. 
Other proposals to eliminate the stability problem of the original Dirac 
model have been approached by considering a spinning bubble in the way studied 
in \cite{chileno} or by the introduction of worldvolume fields pondering the 
spin of the electron \cite{Davidsonspin}. 
Certainly, if a more complete model which includes the tension of the membrane is considered,
we will not necessarily bring an improvement on the dynamical behavior
since it will engender some singularities or discontinuities in the 
potential which can not be ruled out from the outset. This caveat must 
be scrutinized cautiously in the dynamical analysis and provides a 
good motivation for subsequent studies.

Finally, the model described here, results interesting in its own right 
since it has all the hallmarks of cosmological brane models. Among the 
brane cosmology concerns, we can mainly point two of them: their 
gravity and their dynamics. In this context, making use 
of the scheme here developed, it will be interesting to study the cosmology 
associated to the Dvali-Gabadadze-Porrati model, now taking into account 
a term depending linearly on the extrinsic curvature with the purpose 
to remove the pathologies and instabilities of such model~\cite{dgp}. 
An enormous advantage will be that the resultant effective model 
will retain a linear 
dependence on the acceleration of the field variables which open the 
possibility to apply the same quantum strategy followed here.

\section*{Acknowledgements}
We thank an anonymous referee for drawing our attention to the discussion of
some operator order ambiguities in our work and for point us to reference~\cite{Guenkaga}.
This work was partially supported by SNI (Mexico). A.~M.~acknowledges 
financial support from PROMEP (Mexico) under grant UAZ-PTC-086.
E.~R. and R.~C. acknowledge support from CONACYT (Mexico) research Grant 
No. J1-60621-I. E.R was partially supported by the CA-UV: Investigaci\'on 
y Ense\~nanza de la F\'isica. R. C. also acknowledges support from EDI, 
COFAA-IPN and SIP-20100684.

\end{document}